# Emergent superconducting fluctuations in a compressed kagome superconductor

Xikai Wen[1], Fanghang Yu[1], Zhigang Gui[1], Yuqing Zhang[1], Xingyuan Hou[3], Lei Shan[3,4], Tao Wu[1], Ziji Xiang[1], Zhenyu Wang[1], Jianjun Ying[1*], and Xianhui Chen[1,2†]

[1]Department of Physics, and CAS Key Laboratory of Strongly-coupled Quantum Matter Physics, University of Science and Technology of China, Hefei, Anhui 230026, China

[2]CAS Center for Excellence in Quantum Information and Quantum Physics, Hefei, Anhui 230026, China

[3]Information Materials and Intelligent Sensing Laboratory of Anhui Province, Institutes of Physical Science and Information Technology, Anhui University, Hefei 230601, China

[4]Key Laboratory of Structure and Functional Regulation of Hybrid Materials of Ministry of Education, Anhui University, Hefei 230601, China

*E-mail: yingjj@ustc.edu.cn
†E-mail: chenxh@ustc.edu.cn

## ABSTRACT

**The recent discovery of superconductivity (SC) and charge density wave (CDW) in kagome metals $A$V$_3$Sb$_5$ ($A$ = K, Rb, Cs) provides an ideal playground for the study of emergent electronic orders. Application of moderate pressure leads to a two-dome-shaped SC phase regime in CsV$_3$Sb$_5$ accompanied by the destabilizing of CDW phase; such unconventional evolution of SC may involve the pressure-induced formation of a new stripe-like CDW order resembling that in La-214 cuprate superconductors. Nonetheless, the nature of this pressure-tuned SC state and its interplay with the stripe order are yet to be explored. Here, we perform soft point-contact spectroscopy (SPCS) measurements in CsV$_3$Sb$_5$ to investigate the evolution of superconducting order parameter with pressure. Surprisingly, we find that the superconducting gap is significantly enhanced between the two SC domes, at which the zero-resistance temperature is suppressed and the transition is remarkably broadened. Moreover, the temperature dependence of the SC gap in this pressure range severely deviates from the conventional BCS behavior, evidencing for strong Cooper pair phase fluctuations. These findings reveal the** complex intertwining of the stripe-like CDW with SC in the compressed CsV$_3$Sb$_5$, suggesting striking parallel to the cuprate superconductor La$_{2-x}$Ba$_x$CuO$_4$. Our results point to the essential role of charge degree of freedom in the development of intertwining electronic orders, thus provides new constraints for theories.

## 1. Introduction

In strongly correlated systems, most typically the high-temperature superconductors and heavy fermions, the unconventional superconductivity (SC) usually emerges concomitantly with various symmetry breaking orders such as magnetic order, charge density wave (CDW) order, or the electronic nematic order[1-3]. The order parameters of these phases can lead to complex intertwinements[4-6], which could be crucial for resolving the conundrums of pairing mechanisms for the unconventional SC. In the absence of strong electronic correlation, however, the charge/spin orders often act as competitors of SC. Whilst the lack of intricate intertwining orders is in accordance with the mostly conventional SC pairings in the weakly or moderately correlated materials, the recently discovered kagome metals $A$V$_3$Sb$_5$ ($A$=K, Rb, Cs) can probably be a noticeable exception: they hosts a novel charge-ordered state, inside which electronic nematicity and SC emerge with decreasing temperature[7-19]. The enriched electronic orders observed in $A$V$_3$Sb$_5$ provide a unique platform for investigating the unconventional intertwinements[16, 18, 20]; unlike the usual cases for the high-temperature and heavy fermion superconductors where the magnetic fluctuations are believed to play a central role, spin orders are irrelevant for this series, implying that the underlying physics is predominantly controlled by the charge degree of freedom.

Although the SC in $A$V$_3$Sb$_5$ appears to show conventional s-wave characteristics at ambient pressure[21, 22], more fascinating behaviours have been revealed in the $A$ = Cs member with pressure applied[23-26]: the superconducting transition temperature ($T_c$) in CsV$_3$Sb$_5$ displays an intriguing two-dome-like feature under pressure, whereas the onset temperature of the charge order, $T_{CDW}$, monotonically decreases with increasing pressure and is completely suppressed at a critical pressure around 2 GPa[23, 24]. Moreover, a stripe-like CDW order has been revealed between the two SC domes[27], which may be associated with the strong suppression of $T_c$. The unusual evolutions of the SC and CDW orders under pressure, with strong evidence for the interplay between them, are in sharp contrast to the ordinary CDW compounds with conventional SC; they promisingly hint at unusually intertwined electronic orders. Nevertheless, owing to the limitation of experimental probes under high pressure, the nature of the superconducting state and its interplay with the stripe-like CDW order remain largely unknown.

**2. Materials and methods**

2.1. Single crystal growth and characterization

High quality single crystals of $CsV_3Sb_5$ were synthesized from Cs ingot (purity 99.9%), V powder (purity 99.9%) and Sb grains (purity 99.999%) using the self-flux growth method similar to the previous reports [7]. In order to prevent the reaction of Cs with air and water, all the preparation processes were performed in an argon glovebox. After reaction in the furnace, the as-grown $CsV_3Sb_5$ single crystals are stable in the air. The excess flux is removed using water and millimeter-sized single crystal can be obtained.

2.2. Soft point-contact spectroscopy measurements

Soft point-contacts on $CsV_3Sb_5$ were prepared by applying a small drop of silver paint between a 30 $\mu m$ diameter Pt wire with the flat and shiny surface of $CsV_3Sb_5$, which is freshly-cleaved along the c axis. The typical sizes of these contact areas are about 50-150 $\mu m$, including thousands of parallel conducting channels between the sample surface and nanometer-scale silver particles in the paint. G(V) curves were recorded with the conventional lock-in technique in a quasi-four-probe configuration. A DC current bias supplied by Keithley 6221 current source is mixed with ac components from the lock-in amplifier oscillator output, and then applied on the point-contact area. The dc bias voltage was detected by Keithley 2182 nano-voltmeter. The first harmonic response of the lock-in amplifier oscillator input, which is proportional to its point-contact resistance 1/G(V), was collected simultaneously. The measured G(V) curves can be quantitatively analyzed using the well-developed Blonder-Tinkham-Klapwijk (BTK) model [28, 29], with the formular: $G=\int_{-\infty}^{\infty} \frac{df(E-V,T)}{dV}[1 + A(E) - B(E)]dE$, where V is the junction bias voltage, E is the energy, f (E) is the Fermi distribution function, A and B are probability for Andreev reflection and ordinary specular reflection, respectively.

2.3. High pressure measurements

Piston cylinder cell (PCC) was used to generate hydrostatic pressure up to 2.46 GPa, and Daphne 7373 oil was used as pressure transmitting medium. The pressure value in PCC were determined from the superconducting transition of Sn [30]. We used four different soft point-contacts as four probes to measure the resistance. High pressure resistance and pressure-dependent SPCS differential conductance $G(V)$ of $CsV_3Sb_5$ single crystal were measured in a refrigerator system (HelioxVT, Oxford Instruments).

## 3. Results and discussions

In order to investigate the superconducting order parameter of $CsV_3Sb_5$ under pressure, we combine soft point-contact spectroscopy (SPCS) measurement with high pressure technique[31], which makes the identification of the superconducting gap and pairing symmetry accessible[32-34]. We find that the superconducting gap dramatically enhances between the two domes and reaches its maximum value of 1.4 meV around 1.6 GPa, while the $T_c^{zero}$ is strikingly low revealed by simultaneous resistance measurements, having strong violation of BCS theory. In addition, the temperature ($T$) dependence of SC gap severely deviates from the BCS predication, indicating an unconventional superconductivity in this pressure region. Our results show that the unusual competition between the CDW and SC likely involves the phase coherence of the SC state rather than the local pairing amplitude, and support that the superconductivity is strongly correlated with the possible stripe-like CDW order. Such behavior resembles the phase diagram of $La_{2-x}Ba_xCuO_4$[35-39].

We first check the transport properties of the $CsV_3Sb_5$ singly crystal under pressure in piston cylinder cell. Four soft point contacts were used as probes to measure the resistance, while differential conductance of SPCS measurements can be simultaneously recorded on the same crystal, as shown in Fig. 1a. $T$-dependence of the resistance for $CsV_3Sb_5$ under pressure is shown in Fig. 1b. The anomaly in the derivative resistance curves, related to the CDW transition temperature, gradually shifts to lower $T$ with increasing the pressure and disappears at ~2 GPa as shown in Fig. 1b and c. $T_c^{onset}$ and $T_c^{zero}$ can be extracted from the low-$T$ resistance curves (Fig. 1f).

The resistance measurements show a double-dome superconductivity behavior with two maxima located at $P_1 \sim 0.7$ GPa and $P_2 \sim 2$ GPa, suggesting an unusual competition between superconductivity and CDW, which is due to the stabilizing of stripe-like CDW order at high pressure[27]. Remarkably, the superconducting transition is rather broad between 0.7 GPa to 2 GPa, which is consistent with previous results[23, 24]. Alternatively, the superconducting gap closing temperature $T_c^G$ can be obtained from the SPCS curves: The Andreev reflection enhances the conductance in $G(V)$ of the superconducting state relative to the normal state, especially for the zero-bias value $G(0)$; subsequently, $T_c^G$ is defined as the temperature at which the conductance spectrum becomes indistinguishable from that of the normal state. The $G(0)$ curves extracted from the SPCS measurements are plotted as a function of $T$ under various pressures in Fig. S1a and b. One can find that $T_c^G$ is consistent with $T_c^{onset}$ above 0.36 GPa (Fig. S1c). The discrepancy at low pressure can be attributed to the local strain caused by point contacts owing to the surface-sensitive nature of SPCS (for details see supplementary material ).

To hunt for the evolution of the superconducting gap with pressure, we have systematically measured the SPCS differential conductance $G(V)$ under different pressures on a same crystal. The $G(V)$ curves are normalized with the corresponding normal state differential conductance. We then use a generalized isotropic single-gap Blonder-Tinkham-Klapwijk (BTK) model[28, 29] to quantitatively fit the pressure-dependent gap value ($\Delta$). Pressure evolution of the $G(V)$ curves for point contact S1 at 1.5 K and the fitting results (solid black line) are shown in Fig. 2a and b. All the curves show a double-peak feature, and can be well fitted by using the isotropic single-gap BTK model, indicating nodeless s-wave superconducting gap under pressure. We further make a contour plot of $G(V)$ curves under pressure as shown in Fig. 2c. The peak height of $G(V)$ curves shows a clear reduction around $P_1$, then rapidly increases again above 1.3 GPa and becomes almost unchanged at higher pressure as shown in Fig. 2c. Remarkably, the peak-peak distance that is related to the SC gap increases above $P_1$, in stark contrast to the suppression of superconductivity with pressure between $P_1$ and $P_2$, indicating possible unconventional superconductivity in this region. Quantitatively discussion of the extracted $\Delta$ from BTK fitting will come later.

We further measure the $T$ dependence of the $G(V)$ under various pressures as shown in Fig. 3a-g. With increasing $T$, the double peaks gradually shift towards the center, merging into a single zero-bias peak which disappears as $T$ approaches $T_c^{onset}$. The $T$-dependence of $\Delta$ under various pressures can be obtained from the BTK fitting process using a single isotropic gap. The resulting $\Delta(T)$ can be well-described by the conventional BCS gap function $\Delta(T) = \Delta(0) \tanh(1.74\sqrt{T_C/T - 1})$ at 0.36, 1.98 and 2.20 GPa, as shown in Fig. 3h, m and n, respectively. In the pressure range between $P_1$ and $P_2$, however, the gap size shows a highly unusual $T$ dependence which obviously deviate from the conventional BCS model: above $T_c^{zero}$, $\Delta(T)$ shows an almost $T$-linear suppression upon warming before eventually closing at $T_c^{onset}$ (Fig. 3i-l). Such behaviour possibly hints at the unusual interplay of superconductivity and CDW in this region. The $\Delta(T)$ shows weak temperature dependence below 2~3 K, thus the extracted $\Delta$ at 1.5 K in Fig. 2 can be treated as the SC gap at zero temperature.

The amplitudes of superconducting gap $\Delta$ obtained from the BTK model for two point-contacts S1 and S2 are displayed in Fig. 4a. The $\Delta$ extracted from the single-band fitting for the two point-contacts are nearly the same which indicates the reproducibility of our SPCS measurements. Below $P_1$, the gap size gradually increases as pressure is increased, consistent with the $T_c$ enhancement. With further increasing the pressure, $\Delta$ rapidly reaches its maximum value of 1.4 meV with pressure around 1.6 GPa, beyond which $\Delta$ starts to decrease. For conventional superconductors, superconducting gap is expected to be proportional to $T_c$, assuming that the ratio $2\Delta/k_B T_c$ is independent of tuning parameters (for weak-coupling BCS superconductors, $2\Delta/k_B T_c \approx 3.53$). Surprisingly, in our case, $\Delta$ is dramatically enhanced in the region where $T_c$ is suppressed (between $P_1$ and $P_2$), suggesting a strong pressure dependence of the ratio $2\Delta/k_B T_c$ that may come from the unconventional interplay of charge order and SC in $CsV_3Sb_5$.

The abnormal broadening of the superconducting transition between $P_1$ and $P_2$ implies that the $P$-dependence of $2\Delta/k_B T_c$ highly relies on the definition of $T_c$. In Fig. 4b, we present the values of $2\Delta/k_B T_c$ using both $T_c^{zero}$ and $T_c^{onset}$. Above $P_1$, as $T_c^{zero}$ is rapidly suppressed, $2\Delta/k_B T_c$ calculated from $T_c^{zero}$ shows a significant enhancement, reaching an enormous value of ~11 at ~1.5 GPa, which is larger than that in the iron-based superconductors and comparable to the cuprate superconductors[40, 41]. On the other hand, $T_c^{onset}$ exhibits a weaker $P$-dependence between $P_1$ and $P_2$, thus the corresponding $2\Delta/k_B T_c$ yields lower value ~5 (before decreasing to ~4 at $P_2$). In both cases, the value exceeds the BCS expectation of 3.53, placing the superconductivity between P1 and P2 into the strong-coupling regime.

One may assign the broadened superconducting transition and the non-BCS $T$ dependence of the superconducting gap to an inhomogeneous SC state, i.e., a spatial distribution of regions with different $T_c$. Nonetheless, the SPCS spectra obtained on different point-contacts S1-S7 show considerable consistency (see Fig. S2 and S3 in the supplementary material). Moreover, the BCS-like behaviour is recovered (Fig. 3m and n) and superconducting transition becomes sharp again above $P_2$, explicitly ruling out the scenario of pressure-induced inhomogeneity. From the high-pressure NMR work[27], the original CDW order would completely transforms to the pure strip-like charge order phase with pressure between 0.9 to 1.8 GPa. Therefore, the observed unusual superconducting gap in this pressure region can not origin from the mixture phases. In a two-band superconductor, considering the interband coupling, the $T$ dependence of the small gap can also deviate from the BCS-like behaviour[42]. However, the fact that the observed gap already falls into the strong-coupling region and the absence of a larger gap in our SPCS measurements, in principle, exclude this interpretation.

We note that the $T$-$P$ phase diagram of $CsV_3Sb_5$ resembles the phase diagram of $La_{2-x}Ba_xCuO_4$ [35-39](see Fig. S4 in the supplementary material): in the latter system, the formation of a long-range stripe order at the doping level $x = 1/8$ strongly competes with the superconductivity, giving rise to a two-dome-shaped SC regime in the phase diagram; the intertwined SC with the stripe orders (which involves both of charge and spin degrees of freedom) may eventually lead to a pair-density-wave (PDW) order that behaves like a two-dimensional (2D) dynamical SC preceding the realization of three-dimensional bulk SC[35, 43, 44]. The presence of such fluctuating superconducting state could also explain the broadened superconducting transition observed in $CsV_3Sb_5$ between $P_1$ and $P_2$. Moreover, it has been revealed that the $d$-wave SC gap $La_{2-x}Ba_xCuO_4$ is maximized in the vicinity of the stripe-order doping level $x = 1/8$, in spite of the heavily suppressed bulk $T_c$; it also opens at a temperature much higher than the bulk $T_c$ (roughly corresponding to the onset of 2D fluctuating SC) [38, 45]. All these results are qualitatively in agreement with our discoveries in pressurized CsV$_3$Sb$_5$ via SPCS measurements. Hence, it is tempting to attribute the anomalous evolution of SC gap in the intermediate pressure range $P_1 < P < P_2$ (Fig. 4b) to the emergence of a fluctuating SC state[46]: whilst the SC gap remains finite (though no longer follows the BCS model) in the fluctuating range (Fig. 3i-l), the resistivity is non-zero (Fig. 1d) owing to the lack of full SC coherence of the Cooper pairs. The system enters the fluctuating state at $T_c^{onset}$ whereas the coherent bulk SC develops only below a much lower temperature at $T_c^{zero}$ (Fig. 4a).

Such scenario is further corroborated by other experimental evidence. Recent μSR works unveiled low superfluid density in pressurized CsV$_3$Sb$_5$[18, 47] that is comparable with strongly correlated high-$T_c$ superconductors, thus SC phase fluctuations may have a significant influence on the low-temperature properties[46]. More importantly, our NMR studies identify a new stripe-like charge order[27], most likely an analogue of the unidirectional stripe order in cuprates, occurring exactly between the two $T_c$ maxima at $P_1$ and $P_2$. This stripe-like charge order performs as a promising candidate that is intertwined with the SC and consequently gives rise to the novel fluctuating state.

The nature and implication of the fluctuating SC deserve further discussion. Similar fluctuating SC is also observed in the cuprate superconductor Bi$_2$Sr$_2$CaCu$_2$O$_{8+\delta}$, in which the SC gap scales with the onset temperature of SC pairing that is well above $T_c$[40, 48]. In our case, the presence of fluctuating SC in the phase diagram (magenta shaded area in Fig. 4a) coincides with the pressure range in which the stripe-like CDW order develops[27], and $2\Delta/k_B T_c^{onset}$ does not change much under pressure if we assume $T_c^{onset}$ as the onset temperature of SC fluctuation. Hence, the loss of SC coherence, now identified as the reason for the remarkably suppressed $T_c^{zero}$ between $P_1$ and $P_2$, is unambiguously associated with the interplay between charge order and SC. Moreover, the relatively large gap size persisting above $T_c^{zero}$ (Fig. 3i-l) strongly implies that the amplitude of SC order parameter (and also the pairing potential) remains appreciable in the fluctuating state. This immediately point towards a phase fluctuation instead of the conventional Gaussian fluctuation[49] (for the latter, a gap is hardly detectable). Consequently, the non-BCS spectra above $T_c^{zero}$ shown in Fig. 3 are attributed to the Andreev reflection of phase-incoherent Cooper pairs. Both the intertwining between the stripe order and SC and the incoherent SC with strongly fluctuating phase have been established as central puzzles in the research of high-$T_c$ cuprates. Our SPCS results confirm that both can happen in the pressurized $CsV_3Sb_5$, which is distinct from the cuprates in the sense that both strong electron correlation and spin orders are absent herein. The observation of a phase-fluctuating SC state stemming from the intertwined orders involving a stripe-like CDW in such an (at most) moderately correlated system thus provide valuable insights for understanding the universal behaviours and pairing mechanism(s) in unconventional superconductors in a broad perspective.

A further question is whether our results can be related to a PDW order that coexists and/or competes with SC in the pressurized $CsV_3Sb_5$. Though many theoretical works proposed the emergence of PDW in the superconductors with intertwined order parameters[44] and a roton PDW was indeed observed at ambient pressure[20], our SPCS measurements cannot provide evidence for the presence of a second gap (order parameter) that can be assigned to the PDW: the spectra are in accordance with a single gap for which the BCS-type $T$ dependence is destroyed in the phase-fluctuating state (Fig. 3) owing to the weaker contribution to Andreev reflection from the phase-incoherent pairs. The possible existence of PDW is to be verified by further investigations probing the gap function in the momentum space.

**Acknowledgements**

This work was supported by the National Natural Science Foundation of China (Grants No. 11888101 and No. 11534010), the National Key Research and Development Program of the Ministry of Science and Technology of China (No. 2017YFA0303001 and Grants No. 2019YFA0704900), the Anhui Initiative in Quantum Information Technologies (Grant No. AHY160000), the Science Challenge Project of China (Grant No. TZ2016004), the Key Research Program of Frontier Sciences, CAS, China (GrantNo. QYZDYSSWSLH021), the Strategic Priority Research Program of the Chinese Academy of Sciences (Grant No. XDB25000000), the Collaborative

Innovation Program of Hefei Science Center, CAS, (Grant No. 2020HSC-CIP014) and the Fundamental Research Funds for the Central Universities (Grants No. WK3510000011).Innovation Program of Hefei Science Center, CAS, (Grant No. 2020HSC-CIP014) and the Fundamental Research Funds for the Central Universities (Grants No. WK3510000011).

**Author contributions**

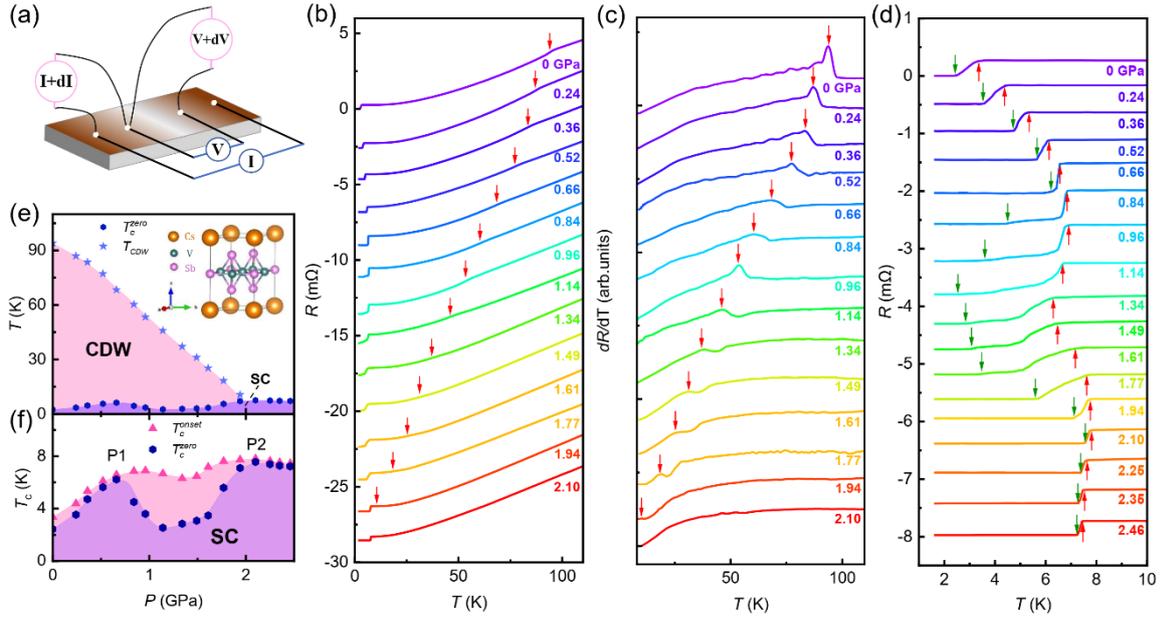

**FIG. 1.** (a) Schematic illustration of soft point contact and resistance measurements on a CsV$_3$Sb$_5$ single crystal. The quasi-four-probe configuration is used for differential conductance measurement. We used four soft point contacts as four probes to measure the resistance. Temperature dependences of resistance (b) and its derivative *dR/dT* (c) under various pressures up to 2.1 GPa. The red arrows indicate the CDW transition temperature $T_{CDW}$. (d) The evolution of superconducting transition temperatures under various pressures. The red and green arrows represent $T_c^{onset}$ and $T_c^{zero}$, respectively. All the curves are shifted vertically for clarity. (e) Pressure-dependent electronic phase diagram of CsV$_3$Sb$_5$. The insert at the upper right corner is the crystal structure for CsV$_3$Sb$_5$. The Cs, V, Sb atoms are presented as brown, cyan and purple balls, respectively. (f) Pressure-dependent superconducting transition temperatures $T_c^{onset}$ and $T_c^{zero}$ are summarized. $T_c^{zero}$ show a double-dome superconducting behavior with two maxima locate at $P_1 \sim 0.7$ GPa and $P_2 \sim 2$ GPa. $T_c^{onset}$ and $T_c^{zero}$ are significant different between $P_1$ and $P_2$, representing rather broad superconducting transition in this pressure region.

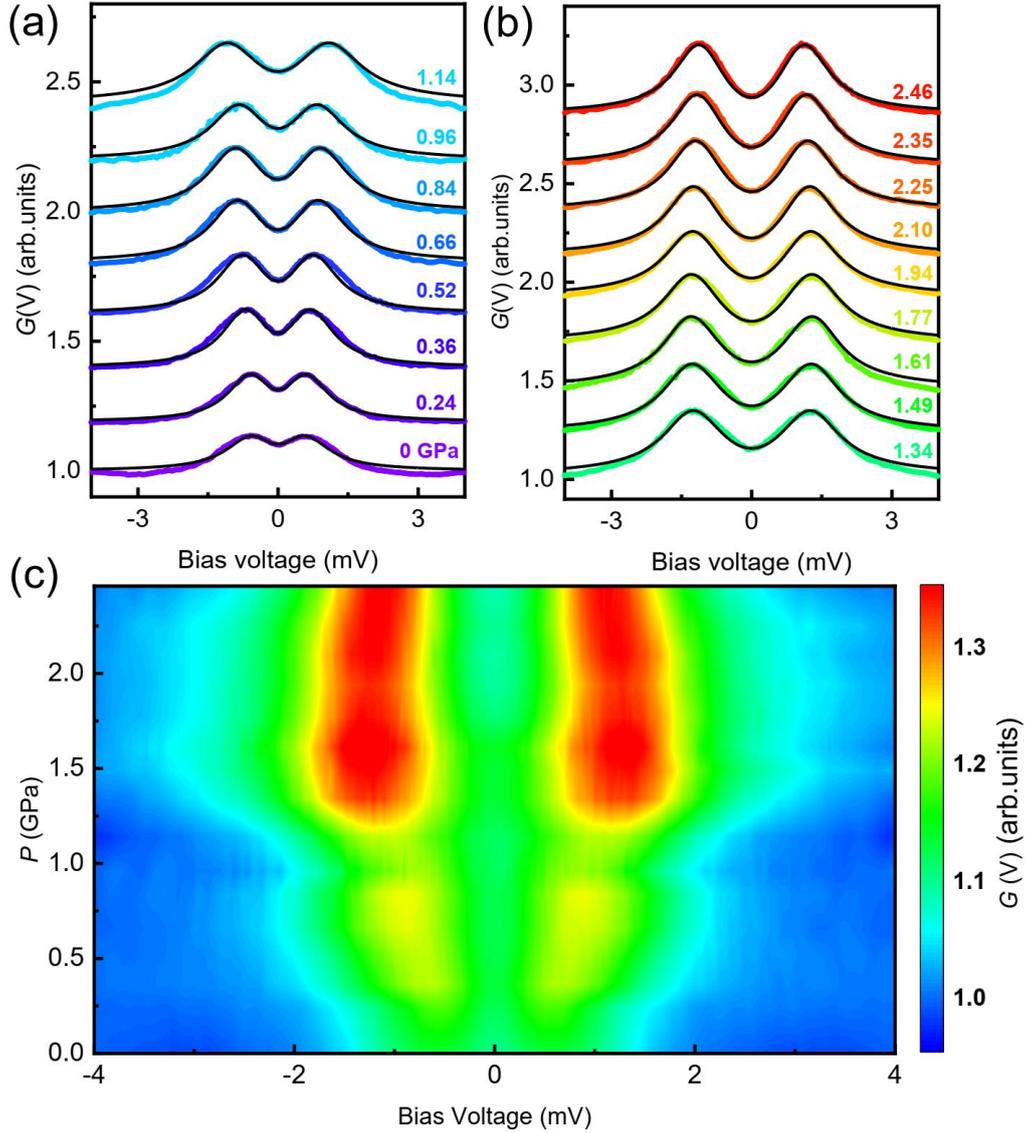

**FIG. 2.** (a), (b) The SPCS differential conductance curves are normalized by dividing the normal state differential conductance. The normalized SPCS differential conductance $G(V)$ for $CsV_3Sb_5$ under various pressures and their fitting results (solid black lines) by using the generalized isotropic single-gap BTK model. All the curves are measured from the point-contact S1 and are shifted vertically for clarity. (c) Contour diagram of the normalized SPCS differential conductance curves under various pressures. The color represents the magnitude of the normalized SPCS differential conductance.

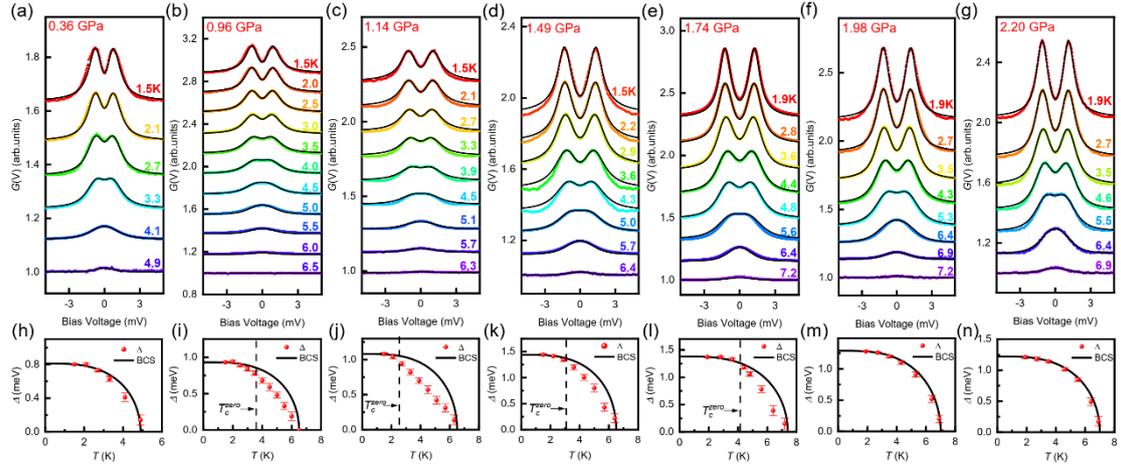

**FIG. 3.** (a)-(g) The temperature dependence of the normalized SPCS differential conductance $G(V)$ under pressures and their BTK fits (solid black line). All curves are shifted vertically for clarity. The $G(V)$ curves with pressure from 0.36 to 1.49 GPa are measured from the point-contact S2, while the other $G(V)$ curves are measured from the point-contact S3 on another crystal. (h)-(n) The temperature-dependent SC gap $\Delta(T)$ (red dots) extracted from the isotropic single-gap BTK fitting. The solid black line is the expected behavior of the empirical BCS gap formula: $\Delta(T)=\Delta(0)\tanh(1.74\sqrt{T_c/T-1})$. The extracted SC gap with pressure between $P_1$ and $P_2$ obviously deviates from the BCS behavior. The deviation temperature is close to $T_c^{zero}$ (black dashed line).

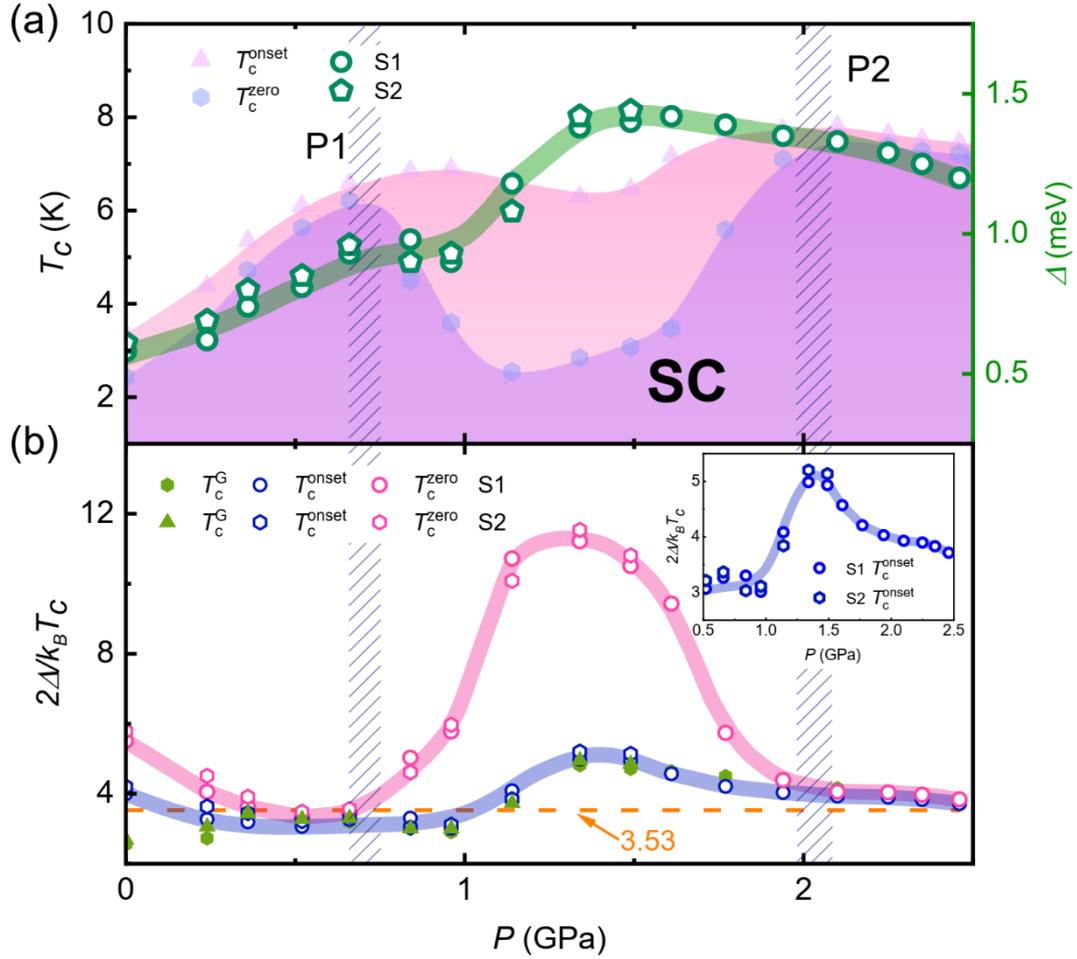

**FIG. 4.** (a) The pressure-dependent of $T_c^{onset}$, $T_c^{zero}$ and SC gap $\Delta$ are summarized. $T_c^{zero}$ shows a double-dome superconducting behavior with two maxima locate at $P_1 \sim 0.7$ GPa and $P_2 \sim 2$ GPa. The pressure-dependent gap $\Delta$ extracted from the BTK model shows a suddenly enhancement above $P_1$, and reaches the maximum value around 1.6 GPa. (b) Pressure dependent of $2\Delta/k_B T_c$, which is calculated by using the different $T_c$ definitions. All the calculated $2\Delta/k_B T_c$ suddenly increase above $P_1$ and keep greater than the weak coupling limit value 3.53 between $P_1$ and $P_2$, indicating the strong coupling superconductivity between $P_1$ and $P_2$.